\documentclass[twocolumn,pra,showpacs]{revtex4-2}
\usepackage{epsfig}
\usepackage{dcolumn}
\usepackage{natbib}
\usepackage[utf8]{inputenc}
\usepackage[english]{babel}
\usepackage{xcolor}
\usepackage{subcaption}
\usepackage{graphicx}
\usepackage{xcolor}
\usepackage{color}
\usepackage{bm}
\usepackage{tabularx}
\usepackage{amssymb}
\usepackage[intlimits]{amsmath}
\usepackage{booktabs}
\usepackage{amsmath}
\usepackage[para,online,flushleft]{threeparttable}

\definecolor{darkcandyapplered}{rgb}{0.64, 0.0, 0.0}
\usepackage[colorlinks=true,allcolors=darkcandyapplered]{hyperref}
\usepackage{physics}
\usepackage{multirow}
\usepackage{float}
\begin{document}
\title{Enhanced nuclear Schiff and electric dipole moments in nuclei with an octupole deformation}

\author{V. V. Flambaum} 
\email{v.flambaum@unsw.edu.au}
\author{A. J. Mansour}
\email{andrew.mansour@student.unsw.edu.au}
\affiliation{School of Physics, University of New South Wales,
Sydney 2052, Australia}
\date{\today}

\begin{abstract}

Deformed nuclei exhibit enhanced moments that violate time-reversal invariance ($T$) and parity ($P$). This paper focuses on the enhanced nuclear electric dipole moment (EDM) and Schiff moment present in nuclei with octupole deformation (pear-shaped nuclei). These moments, which are proportional to the octupole deformation, have a collective nature and are large in the intrinsic frame that rotates with the nucleus. However, in a state with definite angular momentum and parity, $T$ and $P$ conservation forbid their expectation values in the laboratory frame, as nuclear rotation causes them to vanish.

In nuclei with octupole deformation, close opposite-parity rotational states with identical spin are mixed by $T$,$P$-violating nuclear forces. This mixing polarises the nuclear axis along the nuclear spin, allowing moments from the intrinsic frame to manifest in the laboratory frame, provided the nuclear spin $I$ is sufficiently large. 

Using half-life data for $E1$ transitions from the NuDat database, we calculate the intrinsic nuclear EDM $d_{\text{int}}$ for a range of nuclei theorised to exhibit octupole deformation.
    From these values, we independently estimate the intrinsic nuclear Schiff moment $S_{\text{int}}$ and the octupole deformation parameter $\beta_{3}$. Finally, we compare the magnitude of these collective moments in the laboratory frame with the contributions from valence nucleons, providing an estimate of the nuclear EDM and Schiff moment components unrelated to octupole deformation. The uncertainty of our estimates may exceed a factor of 10.
\end{abstract}

\maketitle

\section{Introduction}

Unification theories predict the violation of invariance under the combined operations of spatial inversion ($P$) and charge reversal ($C$), otherwise known as $CP$ symmetry. According to the $CPT$ theorem, $CP$-violation is linked to the violation of time-reversal ($T$).  Therefore,  these theories may be tested via the measurement of time-reversal and parity violating electric dipole moments (EDMs) of particles and atoms. Such measurements have been able to exclude a number of models and significantly reduce the parametric space of other popular models including supersymmetry~\cite{PR,ERK}. An unknown $CP$-violating interaction is also theorised to be the cause of the baryogenesis problem, the matter-antimatter asymmetry in the universe. The magnitude of the atomic EDM is small, making it a priority to search for mechanisms which may enhance it, see e.g.~\cite{Khriplovich1991,Ginges2004,Khriplovich1997}.

The inability of a homogeneous static electric field to accelerate a neutral atom implies that the field seemingly has no effect on the charged nucleus, meaning it must be completely screened by the atomic electrons. Therefore, the nuclear EDM does not interact with the external electric field. One may interpret this as the complete screening  of  the nuclear EDM in neutral atoms and molecules. This was shown to be the case in an arbitrary, non-relativistic system of point-like particles by Schiff in Ref.~\cite{Schiff1963}. However, the electric field inside a finite nucleus does not vanish. To describe this effect, the nuclear ``Schiff'' moment was introduced in Refs.~\cite{Sandars1967,Hinds1980,Sushkov1984,Sushkov1985,Sushkov1986,Ginges2002}. The Schiff moment produced by the proton EDM was calculated in Refs.~\cite{Sandars1967,Hinds1980}. Refs.~\cite{Sushkov1984,Sushkov1985,Sushkov1986} showed that the leading contribution to the nuclear EDM and Schiff moment is the result of $T$,$P$-violating nuclear forces. The effects of both a finite nuclear size and relativistic electrons were calculated in Ref.~\cite{Ginges2002}.

Given the relatively small magnitude of the resultant atomic EDM, it is advantageous to seek the existence of a mechanism which enhances these nuclear moments. Such enhancement is theorised to be present in isotopes with nuclei that exhibit an octupole deformation. These nuclei have a low-lying energy doublet to the ground state, with opposing parity and identical spin. As such, the Schiff moment produced by $T$,$P$-violating forces which admix these states should be enhanced. Nuclei with a soft octupole vibration mode are also theorised to provide a similar enhancement, however the largest enhancement (of the order of $\sim 10^{2} - 10^{3}$ times) is expected in nuclei with an intrinsic octupole deformation, see~\cite{Auerbach1996,Spevak1997}. According to these References, this occurs in specific isotopes of Fr, Rn, Ra and other actinide atoms. Atomic and molecular EDMs which are produced by Schiff moments increase with the nuclear charge $Z$ faster than $Z^{2}$~\cite{Sushkov1984}. This is another contributing factor to the expectation that EDMs in actinide atoms and their molecules are significantly larger than in other systems. Note that the enhancement of $T,P$-violating effects due to the existence of octupole doublets has been noted in Ref.~\cite{Spevak1995}. Earlier such enhancement of $P$-violating effects was noted in Ref.~\cite{Sushkov1980}.

The nuclear EDM and the Schiff moment are proportional to the product of the quadrupole deformation parameter and square of the octupole deformation parameter, $ \beta_{2} \beta_{3}^{2}$~\cite{Spevak1997}. A comprehensive tabulation of the parameters $\beta_{2}$ and $\beta_{3}$ for the ground states of nearly all nuclei can be found in Ref.~\cite{Moller2016}. Estimates of these parameters with alternative nuclear structure models can give differing results, see e.g.~\cite{Ebata2017,Nazarewicz1984}.

A similar enhancement of the nuclear EDM and Schiff moment may also manifest itself in nuclei which are not theorised to have a static octupole deformation, but exhibit a dynamical octupole vibration mode~\cite{Engel2000,Zelevinsky2003,Auerbach2006}. According to Ref.~\cite{Engel2000}, in nuclei with a soft octupole vibration mode, the squared dynamical octupole deformation parameter takes similar values to that predicted for nuclei with a static octupole deformation, $\ev{\beta_{3}^{2}} \sim (0.1)^{2}$, despite the fact that in these nuclei $\ev{\beta_{3}} = 0$. This observation significantly increases the list of nuclei in which we expect nuclear moments to be enhanced.

Refs.~\cite{Auerbach1996,Spevak1997,Engel2003,Dobaczewski2018} have performed numerical calculations of the Schiff moments and estimates of the atomic EDM produced by the electrostatic interaction between electrons and these moments for $^{221}$Fr, $^{223}$Fr, $^{223}$Ra, $^{223}$Rn, $^{225}$Ra, $^{225}$Ac and  $^{229}$Pa. The short lifetime of these nuclei proves to be an issue when it comes to experiments. Such experiments for $^{223}$Rn and $^{225}$Ra have been considered by various experimental groups~\cite{Parker2015,Bishof2016,Tardiff2008}. Despite the Schiff moment enhancement, the $^{225}$Ra EDM measurements presented in Refs.~\cite{Bishof2016,Tardiff2008} have not reached the sensitivity to the $T$,$P$-violating interaction as the Hg EDM experiment~\cite{Graner2016}. These experiments are hindered by the instability of $^{225}$Ra (which has a half-life of 15 days) and a relatively small number of atoms. Refs.~\cite{Flambaum2020,Feldmeier2020} extended the list of candidate nuclei to include stable isotopes $^{153}$Eu, $^{155}$Gd, $^{161}$Dy and $^{163}$Dy, as well as long lifetime nuclei $^{153}$Sm,$^{165}$Er,$^{225}$Ac, $^{227}$Ac, $^{229}$Th, $^{229}$Pa, $^{233}$U, $^{235}$U, $^{237}$Np and $^{239}$Pu.

Using this list of candidates as a guide, we use the half-life data from the NuDat database to estimate the nuclear EDM in the hydrodynamic liquid-drop model, with a particle coupled to a
hydrodynamic core. These values are then used to estimate the nuclear Schiff moment and the octupole deformation parameter. The calculations presented in this paper may serve as a guide for future studies of nuclei theorised to have octupole deformation and may stimulate the undertaking of sophisticated nuclear many-body calculations (as well as atomic, molecular and solid state calculations) for new candidate nuclei. Such many-body calculations are resource-intensive and have thus far been limited to $^{225}$Ra~\cite{Engel2003} (and potentially $^{153}$Eu~\cite{Arvanitaki2024}) over nearly three decades since the original proposal in \cite{Auerbach1996}.

The calculations performed in this paper are also motivated by the work of various experimental groups searching for ultralight dark matter. It was noted in Ref.~\cite{Graham2011} that axion dark matter produces an oscillating neutron EDM, as the axion field is equivalent to a dynamical QCD $\bar{\theta}$-term. The QCD $\theta$-term also produces $T$,$P$-violating nuclear forces, creating $T$,$P$-violating nuclear moments. Correspondingly, the axion field also produces oscillating nuclear $T$,$P$-violating moments~\cite{Stadnik2014}. In order to obtain results for the oscillating $T$,$P$-violating moments, one may replace the $\bar{\theta}$ constant with ${\bar \theta}(t)= a(t)/f_a$, where $f_a$ is the axion decay constant and 

\begin{align}
a = a_{0} \cos (\omega t + \varphi), \ \omega \approx m_{a} \,,
\end{align}

where $\varphi$ is a (position-dependent) phase and $m_{a}$ is the mass of the axion. Assuming that axions saturate the entire dark matter density, the amplitude $a_{0}$ may be expressed in terms of the local dark matter density $\rho_{\text{DM}} \approx 0.4 \ \text{GeV/cm}^{3}$, see e.g. Ref.~\cite{DMdensity},

\begin{align}
a_{0} = \frac{\sqrt{2 \rho_{\text{DM}}}}{m_{a}} \,.
\end{align}  

The effect produced by the oscillating axion-induced Schiff moment of $^{207}$Pb in solid state ferroelectric materials is being probed by the Cosmic Axion Spin Precession Experiment (CASPEr) collaboration~\cite{CasperNew}, while the effect of an oscillating $T$,$P$-violating nuclear polarisability has been measured in Ref.~\cite{Cornell2021} (see theory in Refs.~\cite{pol1,pol2,pol3}). Most recently, the ALP-induced oscillation of the nuclear Schiff moment has been probed in Ref.~\cite{Fan2024}. This experiment was performed using a crystal containing ions of $^{153}$Eu, thereby exploiting the enhancement provided by the octupole mechanism.

\section{Nuclear Shape and Intrinsic Moments}
Let us begin with a review of intrinsic moments caused by a deformation in the nuclear shape, following Ref.~\cite{Spevak1997}. The surface of an axially symmetric deformed nucleus may be described by the expression 

\begin{align}
    R = c_{V}(\beta) R_{0} \left( 1 + \sum_{l = 1}^{\infty} \beta_{l} Y_{l0} \right) \,,
\end{align}
where

\begin{align}
    c_{V} = 1 - \left( \frac{1}{4 \pi} \right) \sum_{l = 1}^{\infty} \beta_{l}^{2} \,,
\end{align}
ensures volume conservation and $R_{0} = r_{0} A^{1/3}$ where $r_{0} \approx 1.1 \ \text{fm}$ is the internucleon distance. In the following calculation, we assume that the coefficient $c_{V} \sim 1$, due to the fact that the nuclear deformations in question are relatively small; $\beta_{2}< 0.2, \beta_{3}, \beta_{4}<0.1$. Reflection asymmetric nuclei require a first order deformation parameter $\beta_{1}$ in order to keep the centre of mass fixed at $z=0$ i.e.,

\begin{align}
    \int z \ d^{3} \vec{r} = 0 \,.
\end{align}
The parameter may be expressed as~\cite{NuclearStructure,Strutinsky1956a,Strutinsky1956b,Dorso1986}

\begin{align}
    \beta_{1} = -3 \sqrt{\frac{3}{4 \pi}} \sum_{l = 2}^{\infty} \frac{(l+1) \beta_{l} \beta_{l+1}}{\sqrt{(2l+1)(2l+3)}} \,.
\end{align}
The Coulomb force dictates that protons and neutrons inside the nucleus are distributed differently. Given proton density $\rho_{p}$ and neutron density $\rho_{n}$, the requirement of a minimum in the energy implies~\cite{Leander1986,Myers1969} 

\begin{align}\label{Coulomb}
    \frac{\rho_{p} (\vec{r}) - \rho_{n} (\vec{r})}{\rho_{p} (\vec{r}) + \rho_{n} (\vec{r})} = - \frac{1}{4C} e V_{\text{Coul}} (\vec{r}) \,,
\end{align}
where $V_{\text{Coul}} (\vec{r})$ is the Coulomb potential created by the distribution of the proton density and $C$ is the volume symmetry-energy coefficient of the liquid drop model. We note that this equality is a kind of static equilibrium
condition.  In the case of octupole vibrations, it may lead to a significant error, especially for oscillations with frequencies of 1 MeV or higher.

To lowest order, the proton density may be written as~\cite{Leander1986}

\begin{align}
    \rho_{p} = \frac{\rho_{0}}{2} - \frac{\rho_{0}}{8} \frac{e^{2} Z}{C R_{0}} \left[ \frac{3}{2} - \frac{1}{2} \left( \frac{r}{R_{0}} \right)^{2} \right. \\ \left.  + \sum_{l = 1}^{\infty} \frac{3}{2l+1} \left( \frac{r}{R_{0}} \right)^{l} \beta_{l} Y_{l0} \right] \,.
\end{align}
where

\begin{align}
    \rho_{0} = \frac{3A}{4 \pi R_{0}^{3}}
\end{align}
The coefficient $C$, whilst not known very accurately, takes values of $\sim$ 20 - 35 MeV in the nuclei of interest~\cite{Leander1986,Strutinsky1956a,Strutinsky1956b,Dorso1986}. An expression for $C$ may yielded from the requirement that 

\begin{align}
    \int \rho (\vec{r}) \ d^{3} \vec{r} = Z \,,
\end{align}
which gives, in lowest order~\cite{Myers1969} 

\begin{align}
    Z = \frac{A}{2} \left( 1 - \frac{3}{10} \frac{e^{2} Z}{C R_{0}} \right) \,.
\end{align}
Due to the relative shift of protons and neutrons inside the nucleus, the intrinsic nuclear frame exhibits a dipole moment given by the expression~\cite{Leander1986,Strutinsky1956a,Strutinsky1956b,Dorso1986,NuclearStructure}

\begin{align} \label{dinttheory}
    d_{\text{int}} = A Z \frac{e^{3}}{C} \frac{3}{40 \pi} \sum_{l = 2}^{\infty} \frac{(l^{2} - 1)(8l+9)}{[ (2l+1)(2l+3)]^{3/2}} \beta_{l} \beta_{l+1} \,.
\end{align}
Further, the intrinsic Schiff moment may be approximated as~\cite{Spevak1997}

\begin{align} \label{SMint}
    S_{\text{int}} \approx e Z R_{0}^{3} \frac{3}{20 \pi} \sum_{l = 2}^{\infty} \frac{(l+1) \beta_{l} \beta_{l+1}}{\sqrt{(2l+1)(2l+3)}} \,.
\end{align}
We note that the expressions presented in this section are based on the hydrodynamic (liquid-drop) model and do not include shell corrections. Refs.~\cite{Leander1986,Butler1991} show that the shell correction to the dipole moment (\ref{dinttheory}) is of the same order of magnitude as $d_{\text{int}}$. The calculations presented in the following sections are only first estimates and may be used in the consideration of future theoretical and experimental studies of enhanced nuclear moments.

\section{Calculation of Enhanced Nuclear Moments} \label{CalculationOfMoments}

The intrinsic nuclear EDM $d_{\text{int}}$ in the nuclei of interest may be calculated from the half-lives of the upper doublet state $t_{1/2}$ found in Ref.~\cite{nudat3}. In general, the half-life of a decaying state may be related to the reduced transition probability between the upper and lower doublet states $B(^{E}_{M}l_{\gamma};J_{i} \rightarrow J_{f})$ using the multipole relation

\begin{align} \label{halflife}
    t_{1/2} = \frac{9 \ln 2}{16 \pi E_{\gamma}^{3} B(^{E}_{M}l_{\gamma};J_{i} \rightarrow J_{f})} \,,
\end{align}
where $E_{\gamma}$ is the energy release during decay and $l_{\gamma}$ is the multipolarity of the transition. Using the rotational model formula, we link the intrinsic electric dipole moment $d_{\text{int}}$ to the reduced electric dipole transition rate

\begin{align} \label{RotationalModelFormula}
    B (E1 ; J_{i} \rightarrow J_{f}) = \frac{3}{4 \pi} d_{\text{int}}^{2} \bra{J_{i} K_{i} 10} \ket{J_{f} K_{f}}^{2} \,,
\end{align}
where $K_{i}$ and $K_{f}$ correspond to the projection of the nuclear spin $I$ onto the rotation axis of the initial and final states respectively. Note that in the case of octupole vibrations, $d_{\text{int}}$ may be regarded as a transition moment (between $0^{-}$ and $1^{-}$ phonon states) rather than an expectation value.

As such, combining Eqs. (\ref{halflife}) and (\ref{RotationalModelFormula}) (considering electrically induced transitions with a multipolarity of $l_{\gamma} = 1$) we yield the following expression for the intrinsic nuclear EDM 

\begin{align}
    d_{\text{int}} = \frac{1}{\bra{J_{i} K_{i} 10} \ket{J_{f} K_{f}}}\sqrt{ \frac{3 \ln 2}{4 E_{\gamma}^{3} t_{1/2}}} \,.
\end{align}
Using half-life data from~\cite{nudat3}, we calculated values of $d_{\text{int}}$ in a range of nuclei which are theorised to have an octupole deformation. Further, substitution of this expression into Equation (\ref{dinttheory}) allows for an independent estimate of the octupole deformation parameter in these nuclei. Considering contributions from the $\beta_{2} \beta_{3}$ term only in Equation (\ref{dinttheory}), we see that

\begin{align}
    \beta_{3} \approx \frac{56 \sqrt{35} \pi d_{\text{int}} C }{9e^{3} A Z \beta_{2}} \,.
\end{align}
Also, taking the ratio of Equations (\ref{dinttheory}) and (\ref{SMint}) (once again considering only the contributions from the $\beta_{2} \beta_{3}$ terms), we yield

\begin{align}
    \frac{d_{\text{int}}}{S_{\text{int}}} & = \frac{5 A \alpha}{14 C R_{0}^{3}} \,,
\end{align}
where $\alpha \approx 1/137$ is the fine structure constant. Thus, using this relation, we may also use our calculations of the intrinsic dipole moment $d_{\text{int}}$ to estimate the intrinsic Schiff moment in the nuclei of interest with a theorised octupole deformation. 

One way in which we identify nuclei that may have an octupole deformation is via an analysis of the rotational spectra, which appears to resemble the spectra of a diatomic molecule of different atoms. We note that the enhancement of the nuclear EDM and Schiff moment in such nuclei is similar to the enhancement of the $T$,$P$-violating effects in polar molecules with non-zero electron angular momentum which have doublets of opposing parity~\cite{Sushkov1978}. In such molecules, the doublet splitting is due to the Coriolis interaction. In nuclei, the splitting is dominated by the ``tunneling” of the octupole bump to the other side of the nucleus, causing a change of the valence nucleon's spin projection to the nuclear axis. In fact, this phenomena may be described as an octupole vibration mode, implying there is no sharp boundary between static deformation in the minimum of the potential energy and a soft octupole vibration when this minimum is very shallow or does not exist. Contrary to the Coriolis splitting in diatomic molecules, the doublet splitting due to the tunneling does not increase with rotational angular momentum - see the nuclear spectra in Ref.~\cite{nudat3}.

This point may require clarification. Static octupole deformation is, in fact, a theoretical construct, referring to a situation where the minimum energy configuration involves an octupole deformation. The potential energy in this case forms a double-well structure with two equal minima (representing a bump on one side and a bump on the other). In a quantum oscillator with such a double potential well, the stationary states are linear combinations of the states where the particle resides in either well. As a result, $\ev{x} = 0$, but $\ev{x^{2}} \neq 0$ and remains relatively large, on the order of the separation between the potential minima. This explains why there is no fundamental difference between "static" octupole deformation and soft octupole vibrations, where $\ev{\beta_3} = 0$ but $\ev{\beta_3^2}$ is large.

However, there is a significant difference in the quantitative approach. When the potential energy does not have a sufficiently deep minimum at a non-zero $\beta_3$, Eq.~(\ref{RotationalModelFormula}) could be applicable under the adiabatic approximation for octupole oscillations, where the rotational frequency is much larger than the octupole vibration frequency. In such cases, we can treat the rotation as occurring for a fixed deformation parameter and then replace the variable $\beta_3^2$ with its average value $\ev{\beta_3^2}$.

In reality however, the rotational frequency is smaller than the octupole vibration frequency. As a result, the use of Eq.~(\ref{RotationalModelFormula}) within our semi-empirical approach provides only an upper limit for the EDM and Schiff moments.

In nuclei with an octupole deformation and nonzero nucleon angular momentum, there exists a doublet of close, opposite parity rotational states $\ket{I^{\pm}}$ with the same angular momentum

\begin{align}
    \ket{I^{\pm}} = \frac{1}{\sqrt{2}} \left( \ket{K} \pm \ket{- K} \right) \,,
\end{align}
where $K$ is once again the projection of $I$ on the nuclear axis. The states of this doublet are mixed by the $T,P$-violating interaction $W$, with mixing coefficient

\begin{align} \label{mixingcoefficient}
    \alpha_{m} = \frac{\mel{I^{-}}{W}{I^{+}}}{E_{+} - E_{-}} \,,
\end{align}
where $E_{+} - E_{-}$ is the energy gap between the states. This mixing polarises the nuclear axis $\vec{n}$ along the direction of the nuclear spin $\vec{I}$
\begin{align} \label{nz}
    \ev{n_{z}}= 2 \alpha_{m} \frac{I_{z}}{I+1} \,.
\end{align}
As such, the intrinsic Schiff moment shows up in the laboratory frame~\cite{Auerbach1996,Spevak1997}

\begin{align} \label{SchiffLab}
    S = 2 \alpha_{m} \frac{I}{I+1} S_{\text{int}} \,.
\end{align}

In a similar way, the intrinsic nuclear EDM $d_{\text{int}}$ of the octupole deformed nuclei of interest also appears in the laboratory frame. Using Equation (\ref{nz}) we obtain

\begin{align}
    d = 2 \alpha_{m} \frac{I}{I+1} d_{\text{int}} \,.
\end{align}
In order to provide analytical estimates of the Schiff Moment/Nuclear EDM in nuclei with an octupole deformation, we require an estimate for the mixing coefficient $\alpha_{m}$. According to Ref~\cite{Spevak1997}, the $T,P$-violating matrix element in the numerator of Equation (\ref{mixingcoefficient}) is approximately equal to

\begin{align} \label{TPmelApprox}
    \mel{I^{-}}{W}{I^{+}} \approx \frac{\beta_{3} \eta}{A^{1/3}} \ \text{eV} \,,
\end{align}
where $\eta$ is the dimensionless strength constant of the nuclear $T,P$-violating potential $W$

\begin{align} \label{TPoddpotential}
    W = \frac{G}{\sqrt{2}} \frac{\eta}{2m} (\sigma \nabla) \rho \,.
\end{align}
Here, $G$ is the Fermi constant, $m$ is the nucleon mass and $\rho$ is the nuclear number density. 

Calculations of the enhanced nuclear EDM and Schiff Moment are shown in Table \ref{FirstTable}. Values for the quadrupole deformation parameter $\beta_{2}$ were taken from Ref.~\cite{Moller2016}. 
The results are presented in terms of the more fundamental QCD $\theta$-term constant $\bar{\theta}$. Within meson exchange theory, the $\pi$-meson exchange gives the dominating contribution to $T$,$P$-violating nuclear forces~\cite{Sushkov1984}. According to Ref.~\cite{Demille2014} the dimensionless strength constant in the $T,P$-violating potential (\ref{TPoddpotential}) may be expressed as (for protons and neutrons)
\begin{align}
\eta_n=-\eta_p =   (-  g \bar{g}_{0} + 5 g \bar{g}_{1} + 2 g  \bar{g}_{2} ) 10^{6}\,, 
\end{align} 
where $g$ is the strong $\pi$-meson - nucleon interaction constant
and $\bar{g}_{0}$, $\bar{g}_{1}$, $\bar{g}_{2}$ are the $\pi$-meson - nucleon $CP$-violating interaction constants in the isotopic channels $T$ = 0, 1, 2. Given this, we may express the results for the nuclear EDM and Schiff moment in terms of more fundamental parameters such as the QCD $\theta$-term constant
$\bar{\theta}$ using the relation $ g \bar{g}_{0}  = 0.21 \ \bar{\theta}$ and $ g \bar{g}_{1} = -0.046 \  \bar{\theta}$~\cite{Demille2014,Bsaisou2015,Yamanaka2017}. Using these results, we obtain

\begin{align} \label{etawrttheta}
 \eta_n=-\eta_p =  4 \times 10^{5} \ \bar{\theta},
\end{align}

We account for the fact that the neutron and proton interaction constants have opposing signs by multiplying by $((N-Z)/N)$, where $N$ is the neutron number of the isotope, taking 

\begin{align}
    \eta = \frac{N-Z}{N}\eta_{n} \,.
\end{align}



Alternatively, the results can be expressed via the quark
chromo-EDMs $\tilde{d}_{u}$ and $\tilde{d}_{d}$: $g \bar{g}_{0}$ = $0.8 \cdot 10^{15}(
\tilde{d}_{u} + \tilde{d}_{d})$/cm, $g \bar{g}_{1}$ = $4 \cdot 10^{15}( \tilde{d}_{u} - \tilde{d}_{d})$/cm~\cite{PR}.

The orders of magnitude of the Schiff moments are consistent with those presented in e.g. Refs.~\cite{Feldmeier2020,Dzuba2020}. Recent studies in Refs.~\cite{Sushkov2024,Arvanitaki2024} have estimated the Schiff moment of $^{153}$Eu using different nuclear models that do not assume a static octupole deformation. The result calculated in Ref.~\cite{Sushkov2024} is two times smaller than our present result,
while the calculations of Ref.~\cite{Arvanitaki2024} predict an even smaller value. Approximations in the evaluation of the $T$,$P$-violating matrix element presented in Eq. (\ref{TPmelApprox}), uncertainties in the input data, the use of a hydrodynamic model and the use of static relations in the case of octupole vibrations, such as that in Eq. (\ref{Coulomb}), lead to an overall uncertainty in our approach that may exceed a factor of 10.

Our calculations of the octupole deformation parameter in the isotopes $^{153}$Gd, $^{161}$Dy, $^{165}$Er, $^{227}$Ac and $^{229}$Pa indicate that is smaller than what is expected for an octupole deformed nucleus, $\beta_{3} \sim 0.1$. As such, we qualify them as candidates for exhibiting a soft octupole vibration mode. 

Calculations were not be able to be performed for $^{163}$Dy, $^{223}$Rn, $^{229}$Th and $^{233,235}$U (which are also theorised to have an octupole deformed nucleus or a soft octupole vibration mode) due to inconclusive $E1$ transition data in Ref.~\cite{nudat3}. For a similar reason, we take the value of $d_{\text{int}}$ for $^{223}$Fr and $^{229}$Pa from other experiments, see Table \ref{FirstTable}.

\begin{table*}[!tbh]
\begin{center}
\begin{tabularx}{\textwidth}{  >{\centering\arraybackslash}X >{\centering\arraybackslash}X >{\centering\arraybackslash}X >{\centering\arraybackslash}X >{\centering\arraybackslash}X >{\centering\arraybackslash}X >{\centering\arraybackslash}X >{\centering\arraybackslash}X }
\hline \hline 
$\multirow{2}{*}{$Z$}$   &  $\multirow{2}{*}{Element}$  & $\multirow{2}{*}{$I$}$ & \multirow{2}{*}{$\lvert d_{\text{int}} \rvert$ ($e$ fm)} & \multirow{2}{*}{$|S_{\text{int}}|$ ($e \ \text{fm}^{3}$)} & \multirow{2}{*}{$\beta_{3}$} & $|d_{\text{lab}}|$ ($ 10^{-2} \ \bar{\theta} \ e$ fm) & $|S_{\text{lab}}|$ ($ \bar{\theta} \ e \ \text{fm}^{3}$) \\
\hline 
    63 &$^{153}$Eu & 5/2 & 0.12 & 9.5 & 0.095 & 0.37 & 0.30 \\
    62 &$^{153}$Sm & 3/2 & 0.13 & 9.8 & 0.095 & 1.0 & 0.75 \\
    64 &$^{155}$Gd & 3/2 & 0.047 & 3.9 & 0.038 & 0.046 & 0.038 \\
    66 &$^{161}$Dy & 5/2 & 0.072 & 5.9 & 0.049 & 0.44 & 0.36 \\
    68 &$^{165}$Er & 5/2 & 0.078 & 6.7 & 0.05 & 0.26 & 0.22 \\
    87 &$^{221}$Fr & 5/2 & 0.14 & 11 & 0.13 & 0.25 & 0.20 \\
    87 &$^{223}$Fr~\footnote{$\lvert d_{\text{int}} \rvert$ obtained from Refs. \cite{KURCEWICZ1992,Sheline1995,Burke1997}.} & 3/2 &  0.038 & 3.0 &  0.028 & 0.019 & 0.015 \\
    89 &$^{225}$Ac & 3/2 & 0.26 & 22 & 0.18 & 3.2 & 2.7 \\
    88 &$^{225}$Ra & 1/2 & 0.29 & 23 & 0.18 & 1.5 & 1.2 \\
    89 &$^{227}$Ac & 3/2 & 0.062 & 5.1 & 0.037 & 0.23 & 0.19 \\
    91 &$^{229}$Pa~\footnote{$\lvert d_{\text{int}} \rvert$ obtained from Ref. \cite{Levon1994}.} & 5/2 & 0.09 & 7.9 & 0.049 & 64 & 56\\
    93 &$^{237}$Np & 5/2 & 0.013 & 1.1 & 0.0054 & 0.0040 &  0.0034 \\
\hline \hline
\end{tabularx}
\caption{Calculations of the intrinsic nuclear EDM from half-life data obtained from Ref.~\cite{nudat3}. Such calculations may be used to estimate the intrinsic Schiff moment $S_{\text{int}}$ and the octupole deformation parameter $\beta_{3}$. 
The theorised octupole deformation of the nuclei in these isotopes implies that it is possible for intrinsic moments to manifest themselves in the laboratory frame due to the mixing of the doublet states by the T,P-odd interaction. As such,
we also present our estimates for the values of the nuclear EDM and Schiff moments in the laboratory frame.}
\label{FirstTable}
\end{center}
\end{table*}


\section{Single-particle estimates of nuclear moments} \label{SingleParticle}

It is instructive to compare the collective contribution due to the octupole deformation presented in the Table \ref{FirstTable} with the single-particle estimates based on the spherical nuclear model~\cite{Sushkov1984}.


Considering a spherical nucleus with one unpaired (valence) nucleon, the nuclear EDM may be expressed as~\cite{Sushkov1984}
\begin{align}
    d_{\text{val}} = - e \left(q - \frac{Z}{A} \right) \xi t_{I} \,,
\end{align}
where $q = 0$ or $1$ for nuclei with an unpaired neutron or proton respectively, $\xi = 2 \times 10^{-21} \eta_{n,p} \ \text{cm}$ and 

\begin{align}
    t_{I} = \begin{cases}
        1\,, \ I = l + 1/2 \,, \\
        - \frac{I}{I+1}\,, I = l - 1/2 \,,
    \end{cases}
\end{align}
with $l$ here being the orbital angular momentum of the unpaired  nucleon, $P = (-1)^{l}$. Using this equation, we estimate the valence nucleon's contribution to the nuclear EDM in Table \ref{SphericalTable}.

\begin{table*}[!tbh]
\begin{center}
\begin{tabularx}{\textwidth}{  >{\centering\arraybackslash}X >{\centering\arraybackslash}X >{\centering\arraybackslash}X >{\centering\arraybackslash}X >{\centering\arraybackslash}X }
\hline \hline 
$Z$   &  Element  & $(I)^{P}$ & $\lvert d_{\text{val}} \rvert$ ($ 10^{-2} \bar{\theta} \ e$ fm)  & $\lvert S_{\text{val}} \rvert$  ($ 10^{-2} \bar{\theta} \ e$ fm$^{3}$)  \\
\hline 
    63 &$^{153}$Eu & (5/2)$^{+}$ & 0.47 & 0.65
    \\
    62 &$^{153}$Sm & (3/2)$^{+}$ &  0.32 & $\sim 0.5$ 
       \\
    64 &$^{155}$Gd & (3/2)$^{-}$ &  0.20 & $\sim 0.5$
     \\
    66 &$^{161}$Dy & (5/2)$^{+}$ &  0.33 & $\sim 0.5$ 
       \\
    66 &$^{163}$Dy & (5/2)$^{-}$ &  0.23 & $\sim 0.5$ 
    \\
    68 &$^{165}$Er & (5/2)$^{-}$ &  0.24 & $\sim 0.5$ 
     \\
    87 &$^{221}$Fr & (5/2)$^{-}$ & 0.35  & 1.7
    \\
    87 &$^{223}$Fr & (3/2)$^{-}$ &  0.49  & 0.59 
    \\
    86 &$^{223}$Rn & (7/2)$^{+}$ &  0.31 & $\sim 0.5$ 
    \\
    89 &$^{225}$Ac & (3/2)$^{-}$ & 0.29  & 1.8 \\
    88 &$^{225}$Ra & (1/2)$^{+}$ &  0.31 & $\sim 0.5$ 
    \\
    89 &$^{227}$Ac & (3/2)$^{-}$ & 0.29  & 1.8
    \\
    90 &$^{229}$Th & (5/2)$^{+}$ &  0.31 & $\sim 0.5$ 
    \\
    91 &$^{229}$Pa & (5/2)$^{+}$ & 0.48  & 0.86
    \\
    92 &$^{233}$U  & (5/2)$^{+}$ &  0.32 & $\sim 0.5$ 
     \\
    92 &$^{235}$U  & (7/2)$^{-}$ &  0.24 & $\sim 0.5$ 
    \\
    93 &$^{237}$Np & (5/2)$^{+}$ & 0.49  & 0.88
    \\
    94 &$^{239}$Pu & (1/2)$^{+}$ &  0.31 & $\sim 0.5$  \\
\hline \hline
\end{tabularx}
\caption{Valence nucleon contribution to the nuclear EDM and Schiff moments.} 
\label{SphericalTable}
\end{center}
\end{table*}

Similarly, the valence nucleon's contribution to the Schiff moment may be expressed as~\cite{Sushkov1984}

\begin{align}
    S_{\text{val}} = - \frac{eq}{10} \xi \left[ \left( t_{I} + \frac{1}{I+1} \right)  \bar{r^{2}} - \frac{5}{3} t_{I} r_{q}^{2} \right] \,,
\end{align}
where $\bar{r^{2}} = \int |\psi|^{2} r^{2} \ dr$ is the mean squared radius of the unpaired nucleon and $r_{q}^{2}$ is the mean squared charge radius of the nucleus. This expression is simplified by noting that it has been experimentally deduced that the magnitude of these radii have close numerical values, meaning we may set $\bar{r^{2}} = r_{q}^{2} = \frac{3}{5} R_{0}^{2}$, following Ref.~\cite{Sushkov1984}. Doing so, we yield the following expression for the Schiff moment

\begin{align}
    S_{\text{val}} \approx - eq \left( t_{I} - \frac{3}{2 (I+1)} \right) A^{2/3} \ 10^{-9} \ \eta \ \text{fm}^{3} \,.
\end{align}
We see that in the case where there exists an unpaired neutron in the nucleus, or an $s_{1/2}$ ($l = 0, I = 1/2$) proton, this expression evaluates to $S_{\text{val}} = 0$. To obtain a non-zero result for $S_{\text{val}}$ in this case, we have to take into account the difference  $\bar{r^{2}} - r_{q}^{2} \sim$ 1 fm$^2$. There exists also a contribution from the polarisation of the nuclear core by the $T,P$-violating field of the valence nucleon~\cite{Sushkov1986}.

From comparison of the results  presented in Table \ref{SphericalTable} and Table \ref{FirstTable}, 
 one observes that the  octupole mechanism  may provide an enhancement of one order of magnitude to the nuclear EDM and an enhancement of three orders of magnitude to the Schiff moment when compared to their values in spherical nuclei 
 \footnote{The calculation for $^{229}$Pa is presented with the caveat that the existence of the very close nuclear energy doublet $\Delta E_{\pm} \approx 60$ eV in $^{229}$Pa is confirmed.}.


Valence nucleon contributions to the nuclear EDM are slightly enhanced in quadrupole deformed nuclei with close opposite parity levels. They are presented in Table 1 of Ref.~\cite{Sushkov1984} (see also \cite{Haxton1983}) and are approximately $|d_{\text{quad,val}}| \sim 0.4 \times 10^{-2} \bar{\theta} \ e \ \text{fm}$. There may also be a small enhancement of the Schiff moments \cite{Sushkov1984}.


\section{Effects of Enhanced nuclear moments in Molecules}

Polar diatomic molecules are another physical system in which close rotational doublet states with opposite parity are evident. Polar molecules exhibit a large intrinsic EDM, meaning they may be polarised by an external electric field. In such molecules, the internal molecular field (which is directed along the axis of the external electric field in a polarised molecule) exceeds the external field by several orders of magnitude. 
For example, the results of Refs.~\cite{Baron2014,Cornell2024} probing the molecules ThO and HfF$^+$ provide limits on the electron EDM which are 2 orders of magnitude stronger than limits placed from the atomic EDM measurement.

Ref.~\cite{Sandars1967} proposed the use of this property to probe the interaction between the nuclear spin and the molecular axis. This occurs due to the interaction of the nuclear Schiff moment with the molecular electrons, and the effect increases faster than $Z^{2}$~\cite{Sushkov1984}. This implies that molecules which contain heavy nuclei  may be advantageous to perform experiments with.

The effective $T$,$P$-violating interaction in molecules is given by

\begin{align} \label{effectivemolecularinteraction}
    W_{T,P} = W_{S} \frac{S}{I} \mathbf{I} \cdot \mathbf{n} \,,
\end{align}
where $\mathbf{I}$ is the nuclear spin, $\mathbf{n}$ is the unit vector along the molecular axis in linear molecules, $S$ is the nuclear Schiff moment and $W_{S}$ is the interaction constant. The constant $W_{S}$ has been calculated for a few structures of interest e.g. TlF~\cite{Petrov2002}, RaO~\cite{Kudashov2013} and PbTi$\text{O}_{3}$~\cite{Titov2016}. While the latter of these structures has been used in the most recent iteration of experiments probing the effect produced by the oscillating axion-induced Schiff moment by the CASPEr collaboration~\cite{CasperNew}, there exist proposals for the use of Eu compounds potentially benefiting from the octupole mechanism, see e.g. Ref.~\cite{ASushkov2023Eu}.

Ref~\cite{Dzuba2020} used the calculations of $W_{S}$ in the structures above, along with their calculations of the atomic EDM in a range of isotopes of interest to estimate the value of $W_{S}$ in the molecules AcF, AcN, AcO$^{+}$, EuN and EuO$^{+}$ containing isotopes of Ac and Eu, which are both theorised to have enhanced nuclear Schiff moments via the octupole mechanism.
Accurate relativistic many-body molecular calculations for actinide and lanthanide molecules have been performed in Ref.~\cite{Skripnikov2020,Oleynichenko2022} (see also the review \cite{Arrowsmith-Kron2024}).
Upon substitution of the Schiff moments calculated in Section \ref{CalculationOfMoments} into Equation (\ref{effectivemolecularinteraction}), an estimate for the energy difference between  states $I_{z} = I$ and $I_{z} = -I$ is obtained. Similar to atoms, molecules containing nuclei which benefit from the octupole mechanism exhibit an enhanced energy difference by $\sim$ 2-3 orders of magnitude.

\section*{Summary}
In this work, we have performed a series of calculations characterising the enhancement of nuclear moments provided by the so called octupole mechanism. Octupole deformed nuclei have a low-lying energy level near the ground state, with opposing parity and identical spin. As such, the $T,P$-odd nuclear moments which are produced through the mixing  of these doublet states by $T$,$P$-violating nuclear forces should be enhanced. Some enhancement may also appear in  nuclei which are theorised to have a soft octupole vibration mode. Therefore, we have estimated  the intrinsic nuclear electric dipole moment in a range of nuclei which are theorised to exhibit  static octupole deformation,  using half-life data for $E1$ transitions. In nuclei with a soft octupole vibration mode our estimates may be used as upper bounds. 

These values were then used to present independent estimates of the octupole deformation parameter of each nucleus, as well as the intrinsic Schiff moment. The mixing of the rotational doublet states in these nuclei polarises the orientation of the nuclear axis along the nuclear spin, resulting in the manifestation of these intrinsic moments in the laboratory frame. This allows us to estimate the enhanced laboratory frame values of the nuclear EDM and Schiff moment in these nuclei. Our calculations indicate an enhancement of approximately 2-3 orders of magnitude of this collective Schiff moment when compared to the valence nucleon  contribution to the nuclear Schiff moment in spherical nuclei. The nuclear EDM may be enhanced up  to one order of magnitude. 

The nuclear EDM contributes to $T,P$-odd effects in atoms and molecules when combined with the hyperfine magnetic interaction between electrons and the nuclear magnetic moment \cite{Schiff1963,Hinds1980,Porsev2011}. However, this contribution increases proportionally with $Z$ \cite{Porsev2011}, while the Schiff moment contribution increases faster than $Z^2$ \cite{Sushkov1984} and dominates in heavy atoms.  
Note that nuclear EDM may, in principle, be measured in an accelerator experiment with highly charged ions.  The ion EDM is $d_{i}=(Z_i/Z) d_N$, where $Z_i$ is the ion charge, $Z$ is the nuclear charge and $d_N$ is the nuclear EDM \cite{AKozlov2012}.


Our calculation of the nuclear EDM and Schiff moment may motivate the undertaking of resource-intensive nuclear many-body calculations for nuclei other than $^{225}$Ra and $^{153}$Eu (as well as corresponding atomic, molecular and solid state calculations)  and has implications on the search for ultralight dark matter which produces an oscillating neutron EDM \cite{Graham2011}, nuclear EDM and Schiff moment \cite{Stadnik2014}. In order to yield enhanced results, experiments measuring the axion field (or axion-like particle field) induced Schiff moment are seeking to probe crystals containing nuclei which benefit from the octupole mechanism. In particular, our calculation of the Schiff moment in $^{153}$Eu may be relevant to experimental groups currently probing crystals containing this isotope such as Refs.~\cite{Fan2024,ASushkov2023Eu}.

\section*{Acknowledgements}
We are grateful to Alex Sushkov, Oleg Sushkov and Chang Zhou for useful discussions. The work was supported by the Australian Research Council Grant No.\ DP230101058.

\bibliographystyle{apsrev4-2}
\bibliography{References.bib}

\end{document}